\documentclass[sigconf,review=false]{acmart}

\AtBeginDocument{%
  \providecommand\BibTeX{{%
    \normalfont B\kern-0.5em{\scshape i\kern-0.25em b}\kern-0.8em\TeX}}}

\copyrightyear{2022}
\acmYear{2022}
\setcopyright{rightsretained}
\acmConference[SEAMS '22]{17th International Symposium on Software
Engineering for Adaptive and Self-Managing Systems}{May 18--23,
2022}{PITTSBURGH, PA, USA}
\acmBooktitle{17th International Symposium on Software Engineering for
Adaptive and Self-Managing Systems (SEAMS '22), May 18--23, 2022,
PITTSBURGH, PA, USA}
\acmDOI{10.1145/3524844.3528062}
\acmISBN{978-1-4503-9305-8/22/05}

\usepackage{ifthen}
\usepackage[normalem]{ulem} 
\newboolean{showcomments}
\setboolean{showcomments}{true} 
\ifthenelse{\boolean{showcomments}}
  {
		\newcommand{\nbb}[2]{
		\fcolorbox{black}{yellow}{\bfseries\scriptsize#1}
		{$\blacktriangleright$\textcolor{blue}{\textit{#2}}$\blacktriangleleft$}
		}
		
		\newcommand{\remarks}[1]{\color{red}[#1]\color{black}}

		\newcommand{\del}[1]{\textcolor{red}{\sout{#1}}} 
		
		\newcommand{\removed}[1]{\cbstart\removedfragile{#1}\cbend{}}
		\newcommand{\removedfragile}[1]{{\color{red}{\sout{#1}}}{}}

  }
  {
		\newcommand{\nbb}[2]{}
		\newcommand{\remarks}[1]{}

		\newcommand{\del}[1]{} 
		
		\newcommand{\removed}[1]{} 
  		\newcommand{\removedfragile}[1]{}

  }
  
\usepackage{setspace} 
\definecolor{keywords}{RGB}{127,0,85}
\usepackage{calc}

\begin{document}

\title{Towards Model Co-evolution Across Self-Adaptation Steps for Combined Safety and Security Analysis}

\author{Thomas Witte}
\author{Raffaela Groner}
\email{<firstname>.<lastname>@uni-ulm.de}
\author{Alexander Raschke}
\author{Matthias Tichy}
\affiliation{%
  \institution{Institute of Software Engineering}
  \city{Ulm University}
  \country{Germany}
}

\author{Irdin Pekaric}
\author{Michael Felderer}
\email{<firstname>.<lastname>@uibk.ac.at}
\affiliation{%
  \institution{Institute of Computer Science}
  \city{University of Innsbruck}
  \country{Austria}
}

\renewcommand{\shortauthors}{Witte and Groner, et al.}

\begin{abstract}
Self-adaptive systems offer several attack surfaces due to the communication via different channels and the different sensors required to observe the environment.
Often, attacks cause safety to be compromised as well, making it necessary to consider these two aspects together.
Furthermore, the approaches currently used for safety and security analysis do not sufficient take into account the intermediate steps of an adaptation.
Current work in this area ignores the fact that a self-adaptive system also reveals possible vulnerabilities (even if only temporarily) during the adaptation.
To address this issue, we propose a modeling approach that takes into account the different relevant aspects of a system, its adaptation process, as well as safety hazards and security attacks.
We present several models that describe different aspects of a self-adaptive system and we outline our idea of how these models can then be combined into an Attack-Fault Tree.
This allows modeling aspects of the system on different levels of abstraction and co-evolve the models using transformations according to the adaptation of the system.
Finally, analyses can then be performed as usual on the resulting Attack-Fault Tree.
\end{abstract}




\maketitle

\section{Introduction}\label{sec:introduction}
The use of self-adaptive systems is steadily increasing, especially, e.g, in the context of cyber-physical systems or autonomous vehicles.
The advantage of such systems is that they are aware of their state and the state of their environment and, if necessary, decide for themselves whether to adapt when a state changes, and if so, how. 
However, this advantage also poses serious risks, since classical analyses are no longer applicable due to their inability to take the adaptation into account.
Since a faulty system can cause damage to property, health and the environment, a consideration of the safety of self-adaptive systems is essential.
But this is still not sufficient, since faulty system behavior can not only occur due to errors in the system, but can also be deliberately provoked by attacks.

Current research in the field of combining safety and security for self-adaptive systems is not sufficient. 
There is some existing work in this field, for example by Fovino et al.~\cite{NAIFOVINO20091394} or Kumar and Stoelinga~\cite{7911867}, that combine Fault Trees and Attack Trees to Attack-Fault Trees.
The Attack-Fault Trees presented are not specialized to the challenges of self-adaptive systems, since they are created manually and not updated automatically according to an adaptation performed.
\v{C}au\v{s}evi\'{c} et al.~\cite{8753949} present in their vision paper their idea for combining safety and security for self-adaptive systems using different models at design and at runtime.
Another example is the work of Lui et al.~\cite{10.1145/3055186.3055193}, that presents an approach of defining safety and security goals and then determining which attacks could compromise them.
These works are a first starting point, but both, \v{C}au\v{s}evi\'{c} et al.~\cite{8753949} and Lui et al.~\cite{10.1145/3055186.3055193}, lack the consideration of the safety and security risks arising during an adaptation. 
The fact that an adaptation is not performed in zero time and cannot be considered as an atomic transaction is not taken into account.

To close this gap, we propose a modeling approach, which consists of several models, that describe different aspects of a system.
We present a model to describe the data flow between individual components and a model to describe the deployment of the components of a system.
These two models are closely representing the system and are able to adapt and co-evolve with the self-adaptive system.
To model the individual steps of an adaptation, we use transformations which, when applied to the models, lead to the individual transient states of an adaptation process.
The safety aspects are modeled using Fault Trees~\cite{vesely1981fault} and the security aspects are modeled using Attack Trees~\cite{schneier2015secrets}.
Since Fault Trees are modeled mainly at the component level and Attack Trees at a more detailed level, e.g., at the protocol level, these two models cannot be easily combined.
Therefore, we outline our idea on how we can combine the information from our models to generate an Attack-Fault Tree~(AFT)~\cite{7911867}.
The resulting AFT then serves as the basis for the associated safety and security analyses.

In the following section, we discuss related work and in Section~\ref{sec:proposedApproach} we present our proposed modeling approach using an illustrative example.
In Section~\ref{sec:analysis}, we provide an outlook on the planned analyses, which are based on our modeling approach. 
Finally, in Section~\ref{sec:conclusion} we summarize and conclude this paper.



\section{Related Work}\label{sec:relatedWork}
In this section we relate our research to other existing approaches. In the context of SAS several work regarding security aspects 
were published. One large topic in this area is the automatic detection of attacks by establishing intrusion detection systems (IDS), e.\,g\ \cite{IDS1,IDS2}. The authors of \cite{ACSThreatDatabase} mention a "threat database" that is used for automatic adaptation of Android Application's permissions, but in contrast to this work, this database is fed manually and not by mining CVS notifications. Pianini et al. "look at security in the context of CASs [collective adaptive systems] by reasoning on how existing patterns and recommendations may apply under the assumptions and characteristics of such systems." \cite{Pianini2019}. According to their taxonomy our work covers all of the four layers presented in the context of the "Defense in depth" principle: application, middleware, platform, and close-to-metal. 
While the mentioned research focus on security, there exist also some work regarding safety of SASs (e.\,g.\ \cite{Scandurra2021, Klumpp2016}). Only some researchers consider safety and security aspects together as shown in the following.

In \cite{10.1145/3055186.3055193} the authors present a safety-security co-design engineering process that is tailored for platooning system. In this process the security goals are derived from the safety goals and an attack model. Similarly, in \cite{S3} the usage of digital twins is suggested to support the identification of safety and security goals. Whereas these approaches focus on the identification of abstract safety and security problems, the vision described in \cite{8753949} introduces some ideas how these models could be used. The authors mention architecture and behavioral models that shall be used to analyze safety and security requirements at run-time of a self-adaptive system. However, no details are given, how these models should look like. In our work we introduce concrete architecture and deployment descriptions that allow for an integration of known or arising vulnerabilities of used components into an analysis framework at run-time, thus bridging the gap between abstract high-level safety goal descriptions and the low-level distributed execution of a system on different platforms.

The approach of Khakpour et al. \cite{Khakpour2019} focuses on vulnerabilities of the system in transient states that occur especially during architecture-based adaptations. They manually add  vulnerability information to the architecture description by utilizing Acme's properties of components. Finally, they use MulVAL\cite{MulVAL2005} to generate a probabilistic attack graph. We also take the adaptation process into account by modeling the adaptation as graph transformation (similarly to Bucchiarone et al. \cite{Bucchiarone2015}) consisting of different steps. We extend this approach by considering security with respect to safety aspects. Instead of defining vulnerabilities manually, we mine them for the components involved from CVE (Common Vulnerabilities and Exposures) databases and the architecture description. Furthermore, we model not only components but also protocols or hardware in our architecture description. The generation of attack trees (in particular for a network-based model) was also done by Kotenko et al.\cite{6568374} and Ou et al. \cite{Ou2006}. 

The approach presented by Priesterjahn et al. \cite{Priesterjahn2013} analyzes the time needed to finish a system adaptation before a hazard occurs. The authors introduce min-max execution time intervals for each component in order to analyze the propagation time of a failure through the system, but they do not take security aspects into account.

Several approaches influenced our ideas how to combine the different models: 
The combination of fault trees with attack trees can be done by introducing new gates and symbols as suggested by Kumar and Stoelinga in \cite{7911867} or Stoelinga et al. in \cite{Stoelinga2}. Another alternative is a less tight coupling just by allowing roots of attack trees as basic events in fault trees. We adopted this suggestion introduced by Steiner and Liggesmayer \cite{SteinerLiggesmeyer2016} and Fovino et al. \cite{Fovino2009}. 

Finally, the comparison of used libraries to CVE database entries is realized by commercial products like snyk\footnote{\url{https://snyk.io}} and for Java by Viertel et al. \cite{Viertel2019}.


\section{Proposed Approach}\label{sec:proposedApproach}

Safety and security properties are intertwined and must not be analyzed in separation. In addition, the resulting model must consider possible architecture adaptations of the system (which have an impact on these safety and security properties) and allow the model to adapt accordingly and automatically.

\begin{figure}[htbp]
    \centering
    \includegraphics[width=\linewidth]{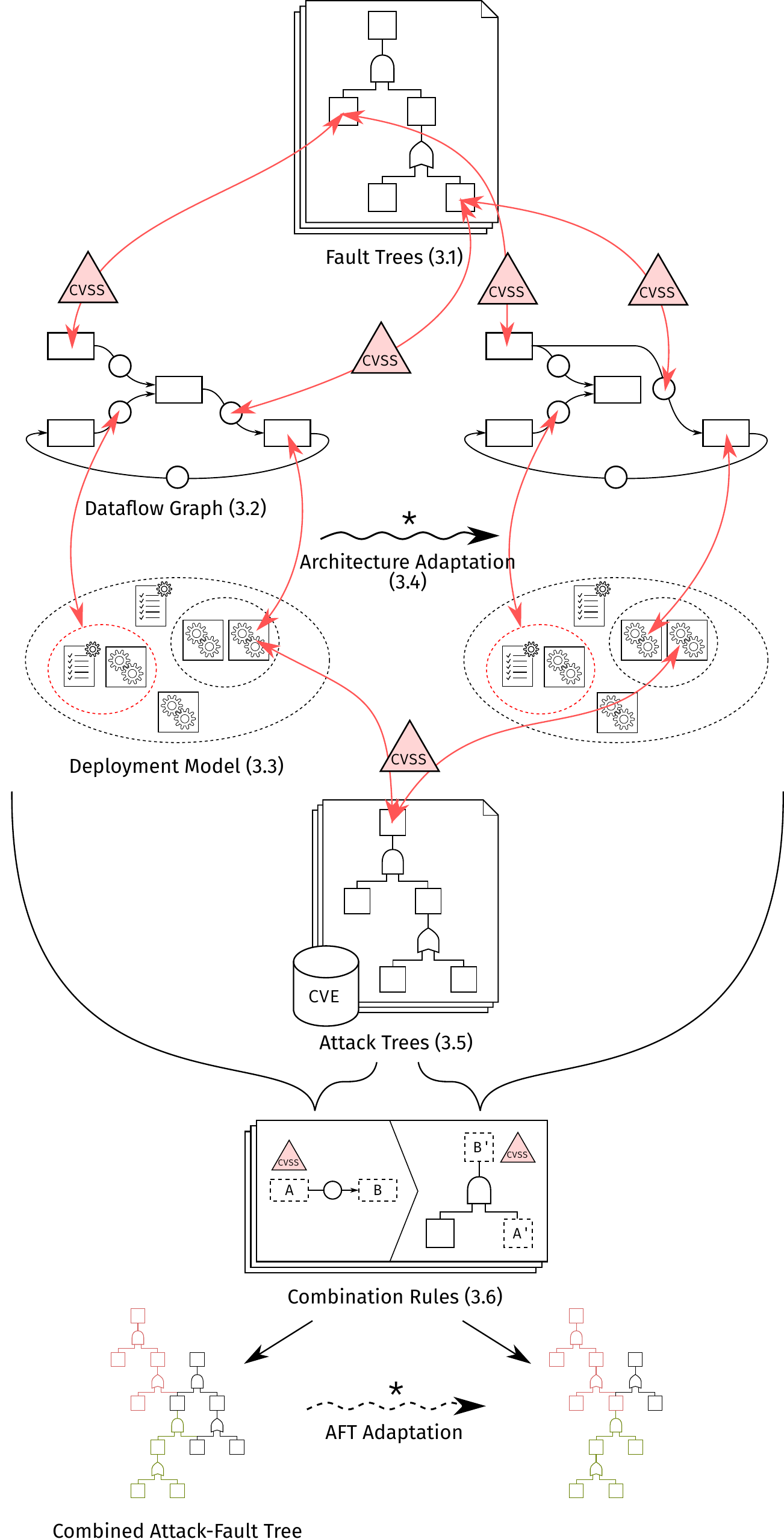}
    \caption{Overview of the modeling approach.}
    \label{fig:overview}
\end{figure}

Figure~\ref{fig:overview} gives a broad overview over our proposed modeling approach for self-adaptive systems. Sections~\ref{sec:fault_trees} and \ref{sec:attack_trees} describe how Fault and Attack Trees can be modeled independently by experts or mined from vulnerability databases. Section~\ref{sec:dataflow} introduces the logical Dataflow Graph that can be tied to events in these Fault or Attack Trees. These links may be annotated with additional impact requirements--minimum CVSS~(Common Vulnerability Scoring System) impact scores required to trigger the linked event. In order to allow attacks to reference specific libraries, protocols or hardware, a Deployment Model is introduced in Section~\ref{sec:deployment}. Graph transformation rules are used to describe the system adaptation (Section~\ref{sec:reconfiguration}). From these distinct models, a combined AFT is generated by using general combination rules, representing common weaknesses and attack patterns (Section~\ref{sec:combination}). Analysis (Section~\ref{sec:analysis}) is then performed in two dimensions: state space exploration of the transformation steps and stochastic model checking on the combined AFTs.

In order to illustrate our proposed approach, we use a small, simplified example, due to space limitations.
Our example consists of a ground control station (GCS) that communicates with an unmanned aerial vehicle (UAV), i.e. quadcopter, during a given mission via WiFi.
The GCS sends flight commands to the UAV and receives telemetry data.
If the UAV passes areas where the WiFi signal is too weak, the adaptation process starts and the system switches to longer range radio communication.
The UAV then receives the command to fly back closer to the GCS.
If the UAV is close enough to the GCS again, the system completes its adaptation by switching back to communicate via WiFi.

%
%

\subsection{Fault Trees}
\label{sec:fault_trees}
In order to model possible safety hazards, we use \emph{Fault Trees} (FT).
Starting from an undesired event and taking into account the system and its environment, it is analyzed which events lead to the occurrence of the undesired event analyzed~\cite{vesely1981fault}.

\begin{figure}[htbp]
    \centering
    \includegraphics[width=0.75\linewidth]{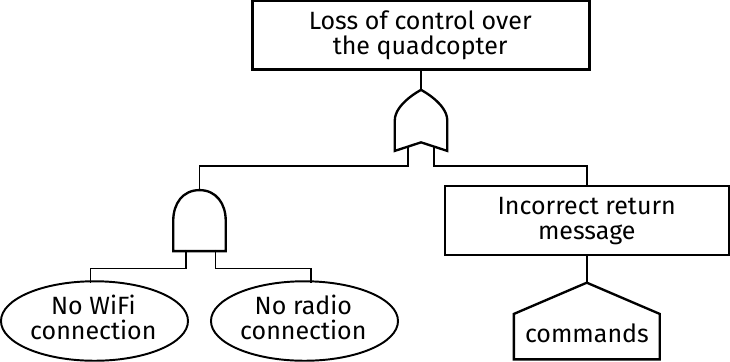}
    \caption{Excerpt of a Fault Tree for the example.}
    \label{fig:faultTree}
    \vspace{-0.5cm}
\end{figure}

There are several variants of Fault Trees~\cite{RUIJTERS201529}, we use a simple one as a starting point, because it can be easily extended if necessary.
The leaves of a Fault Tree are so-called basic events that do not depend on other events and do not require further specification.
The inner vertexes of a Fault Tree are so-called intermediate events, which are events that depend on other events.
Gates are used to express the dependencies between events by linking them logically, e.g. by AND or OR~\cite{vesely1981fault}.
Basic events in a Fault Tree can link to elements of the Dataflow Graph (cf. Section~\ref{sec:dataflow}). We introduce an additional element, which is modeled as an external event (house shape in Figure~\ref{fig:faultTree}) that contains the name of the referenced Dataflow Graph element and the minimum required CVSS impact on the dataflow element to trigger the linked event.
The impact specification corresponds to the CVSS metrics: for example, an event can cause or require an impact on the Confidentiality, Integrity or Availability of the channel or component. The basic event "connection lost" in a Fault Tree might reference one or more dataflow channels and require a high impact on the availability (e.g. caused by a DoS-attack) in order to activate this basic event. Vice versa, the top event "data corruption in component" of an Attack or Fault Tree can reference a component of the \emph{Dataflow Graph} and cause a high integrity impact on the component if the event is activated.

In our approach, we assume that a Fault Tree is modeled for each fault manually by a safety expert, resulting in a forest of fault trees. These Fault Trees are immutable, however, the linked dataflow elements or their deployment might change during reconfiguration.

Figure~\ref{fig:faultTree} displays an excerpt of a Fault Tree for our UAV example.
The analyzed safety risk is the loss of control of the UAV, which can lead to theft of the UAV, property damage or even injuries.
One way that this undesirable event occurs is through the occurrence of the two basic events (shown as ellipses in Figure~\ref{fig:faultTree}) “No WiFi connection” AND “No radio connection”. 
The intermediate event "Incorrect return message" is linked to the "commands"-channel in the Dataflow Graph using an external event. 
Since there are other combinations of events that lead to loss of control, the AND-gate and the intermediate event are connected to the top event using an OR-gate. 
For reasons of readability, we have omitted further parts of the Fault Tree.


\subsection{Dataflow Graph}
\label{sec:dataflow}

The dataflow between the software components of an application, subsystems and physical components contains information how failures or attacks may spread and proliferate. A \emph{Dataflow Graph} that contains the system's components and abstract connections, i.e. not regarding the deployment or realization of the data channel, is used to determine whether an event in one part of the system can have an impact on other parts of the system.

The high level of abstraction of this \emph{Dataflow Graph} simplifies modeling implicit or indirect information flow, e.g. side channels that may leak information or sensors measuring the system state through the environment. 
This logical dataflow is intentionally architecture and technology agnostic to possibly model any kind of system. The specific protocols or technology used is specified in the \emph{Deployment Model}.

The \emph{Dataflow Graph} consists of \emph{components} and \emph{channels} between these components. Components can send messages through directed channels with an arbitrary number of senders and receivers per channel. 

A fault or attack event (in the respective tree) can reference a component or channel that is affected by or the source of the event. The relation of linked Fault or attack events, e.g. that an attack event has a cascading effect affecting other components, is expressed through combination rules (Section~\ref{sec:combination}) that define the structure of the generated AFT.


\begin{figure}[htbp]
    \centering
    \includegraphics[width=0.85\linewidth]{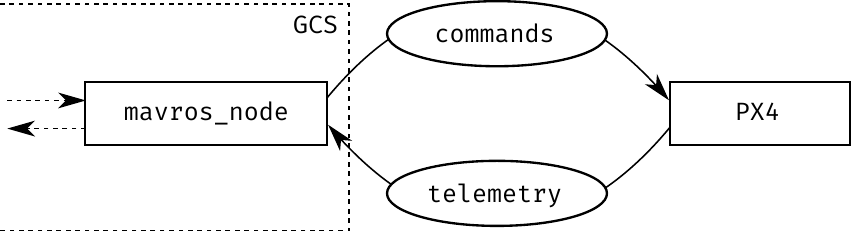}
    \caption{Excerpt of the Dataflow Graph for the example.}
    \label{fig:dataflow_example}
    \vspace{-0.3cm}
\end{figure}

Figure~\ref{fig:dataflow_example} shows an excerpt of the Dataflow Graph for our running example. The mavros\_node in the GCS communicates with the quadcopter (PX4) through a command and telemetry channel. The Fault Tree can reference the \emph{commands} through an external event irrespective of the realization of the channel (WiFi or radio).

The Dataflow Graph can be generated if the platform or framework used offers appropriate introspection capabilities. For example, the node graph of a running ROS~\cite{quigley2009ros} system can be obtained at runtime and directly transformed into a Dataflow Graph. Alternatively, the Dataflow Graph can be derived statically from e.g. SysML internal block diagrams. Additional dataflow information, implicit or indirect channels can be added manually ahead of time.


\subsection{Deployment Model}
\label{sec:deployment}
Every component in the Dataflow Graph runs on a specific hardware (HW) with a given operating system (OS) that provides a set of libraries that might be used by the components. Similarly, every communication channel can use a protocol (stack) that is typcially also implemented in libraries running on an OS on a given HW. Each involved unit (protocol, library, HW) can have one or more vulnerabilites that could be exploited to attack the system. 

The goal of the Deployment Model is on the one side to model the overall architecture of a system on different levels of abstraction and on the other side to define the deployment of the components and channels of the Dataflow Graph on the different parts of the system. 

For this purpose, we define a transitive dependency relation $c \rightarrow D$ with $c$ is the identifier (e.\,g.\ a common platform enumeration entry (CPE)\footnote{a unified naming scheme for hardware, software, and packages, see \url{https://nvd.nist.gov/products}}) of a component/channel/protocol/library/set of libraries/hardware/proxy, $D$ is a set of protocols/libraries/hardware components, and $c$ depends on the entries given in $D$. Types are added to each of these elements for a better differentiation. The combination of dependencies describes the set of platforms on which components can be deployed. In this context, a \textit{proxy} is an abstract placeholder for a concrete instance of a platform. In the following example, \textit{WiFi} is a proxy for a concrete instance of an existing WiFi network with certain properties. The instantiation of this proxy is done in a separate entry.  

The deployment of a component (of the Dataflow Graph) is just a top level dependency of the Deployment Model. Thus, the highest level of the Deployment Model correspond to elements of the Dataflow Graph and the lowest level correspond to (potential) vulnerable elements like protocols, libraries, and hardware components. An architecture reconfiguration can be modeled as a change of the deployment of a component which includes the exchange of a component by a new one. 

In our running example, an excerpt of a simplified Deployment Model might look like this (for sake of brevity, only left-hand-side types are shown):

\noindent 
$\textit{PX4:Component} \rightarrow \{\textit{Mavlink2.0}, \textit{pixhawk, ...}\}$\\
$\textit{Mavlink2.0:Protocol} \rightarrow \{\textit{MavlinkLib, UDP, UART, ...}\}$\\
$\textit{commands:Channel} \rightarrow \{\textit{Mavlink2.0, WiFi}\}$\\
$\textit{WiFi:Platform} \rightarrow \{\textit{IEEE 802.11n, WPA2, UDP, TCP/IP, ...}\}$\\
$\textit{Radio:Platform} \rightarrow \{\textit{UART}\}$ \quad $\ldots$

Note, that some of the dependencies can be derived automatically (e.g. by examining the dependencies of a linux binary or by retrieving the installed library versions of an OS).

Due to the transitive property of the relation it is possible to combine the dependency sets, e.g.: $\textit{commands} \rightarrow \{\textit{Mavlink2.0,}$
$\textit{MavlinkLib, IEEE 802.11n, WPA2, UDP, TCP/IP, UART, ...}\}$. 

\subsection{Architecture Adaptation}
\label{sec:reconfiguration}
In order to describe the adaptation of a system, we use graph transformations.
Thereby, the individual steps of an adaptation to be performed are described by several individual transformation rules.
These transformation rules can then be applied to the Dataflow Graph and the Deployment Model to automatically achieve the transient states during an adaptation.

We use Henshin~\cite{10.1007/978-3-319-61470-0_12} to define the transformations. 
This is a declarative transformation language where transformation rules are created by defining a precondition that a model must satisfy in order for the rule to be applied.
If the precondition is satisfied, the model is updated to satisfy the postcondition of the rule.

The first step of the adaption performed in our UAV example is the switch between the communication via WiFi to radio.
This can be described in the textual syntax of Henshin with the following excerpt:\\
\textbf{\color{keywords}{graph}} \{ \\
\indent \textbf{\color{keywords}{node}} commands:Channel\\
\indent \textbf{\color{keywords}{node}} WiFi:Platform\\
\indent \textbf{\color{keywords}{node}} Radio:Platform\\
%
\indent \textbf{\color{keywords}{edges}}[(\textbf{\color{keywords}{delete}} commands->WiFi:depends),\\
\indent \indent \indent \indent (\textbf{\color{keywords}{create}} commands->Radio:depends)]\}


\begin{spacing}{1.4}

\end{spacing}
The precondition specifies that the \emph{commands}-channel depends on \emph{WiFi}.
The postcondition states that the \emph{commands}-channel should depend on \emph{Radio} and no longer on \emph{WiFi}. 
This change is defined in Henshin by the keywords delete and create. 
Where delete marks the parts of the precondition that must not exist after the application. 
And create marks the parts of the postcondition that will be created after the application. 
Parts without one of the two keywords are part of the precondition and the postcondition, and must be valid before and after the transformation application

In order to model the entire adaptation process, further transformations must be defined manually, which describe also the individual transient adaptation steps.
In our example, this means, e.g, that a transformation must be defined, which switches the communication back again, in order to describe the entire adaptation.




\subsection{Attack Trees}
\label{sec:attack_trees}

Potential attacks on platforms and protocols are modeled using \emph{Attack Trees} (AT).  An Attack Tree represents various ways and necessary steps through which an adversary can attack the system.
Due to possible system reconfiguration, it is necessary to generate Attack Trees for every platform, library or protocol used in the system. The target of such an Attack Tree (e.g. a library) is modeled as an external event that can be linked to the respective resource in the Deployment Model.

Attack Trees are generated using a semi-automated approach, similar to \cite{Ou2006} by incorporating the data from three sources: CVE, CVSS, and system library data. Each attack targets a platform, identified by its CPE entry, and may use one or more vulnerabilities to impact it. The vulnerability data is obtained by implementing active monitors allowing the collection of the most recently published CVE data, the generation of corresponding Attack Trees and creation of an attack database for the system. The characteristics and severity of vulnerabilities are represented using CVSS data. Finally, the system library data includes all the libraries and applications, as well as their specific versions that are used by a self-adaptive system. These can be linked to the CVE data because it also includes information regarding the library that was affected by a specific vulnerability.

\begin{figure}[htbp]
    \centering
    \includegraphics[width=0.65\linewidth]{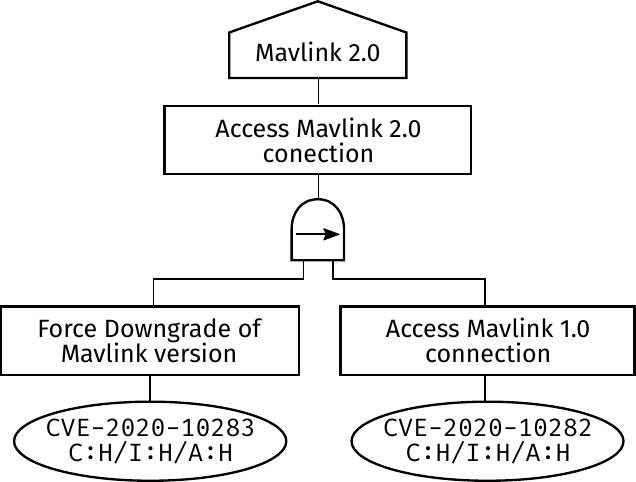}
    \caption{Example of an attack tree.}
    \label{fig:attackt}
\end{figure}

The root element (external event) describes the target of an attack and serves as an interface to link the attack to the combined AFT. For example, the exploitation of two vulnerabilities could result in an attack on the \emph{Mavlink} Protocol (Figure \ref{fig:attackt}). 
Multiple vulnerabilities (CVEs) can be linked together to build an attack path. In addition, the CVSS:3.1 vector is added to the CVE identifier in order to provide additional vulnerability information such as exploitability, impact, temporal score, and environmental score metrics for each basic attack step. This also includes impact measures on CIA metrics. By utilizing the vector data, we limit applicable combination rules to rules matching the characteristics of the attack. 

\subsection{Model Combination}
\label{sec:combination}

Joint analysis of safety and security requires combining Attack and Fault Trees into AFTs using additional information from the Dataflow Graph and Deployment Model. Attack and Fault Trees often operate on different levels of abstraction, i.e. attacks mostly target libraries, frameworks, protocols or hardware while basic events of Fault Trees are mostly triggered by missing or corrupted logical data flow or components. This different levels of abstraction are mirrored in the logical Dataflow Graph and the Deployment Model which specifies concrete systems and software implementing this data flow.

\begin{figure}[htbp]
    \centering
    \includegraphics[width=0.8\linewidth]{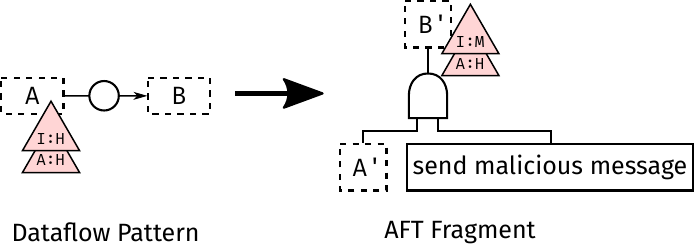}
    \caption{Example combination rule for CWE-20: Improper Input Validation.}
    \label{fig:combination_rule}
    \vspace{-0.3cm}
\end{figure}

Using a rule based translation scheme, patterns in the data flow or deployment model create partial AFTs -- representing generic attack patterns or common weaknesses -- that are connected to external events in the Attack and Fault Trees, combining them and bridging the abstraction gap. Rules are derived from Common Weakness Enumeration (CWE) as well as Common Attack Pattern Enumeration and Classification (CAPEC) data. These combination rules can, for example, translate a logical data flow between two components $A$ and $B$ into a tree fragment shown in Figure~\ref{fig:combination_rule}. Any Attack or Fault Tree referencing component $A$ or $B$ can be connected to the events $A'$ or $B'$ respectively. Such a rule might represent the additional attack step needed to forge an invalid message in component $A$ and sending it in order to compromise component $B$. To forge such a manipulated message, the attack or fault event connecting to $A'$ needs to satisfy additional impact criteria, in this case on the \emph{integrity} or \emph{availability} of component $A$. These impact requirements, i.e. required values in the CVSS vector, serve as an interface to combine only compatible events. Combination rules, fault or attack events as well as references to elements in other models can be annotated with such impact specifications.

\begin{figure}
    \centering
    \includegraphics[width=\linewidth-0.5cm]{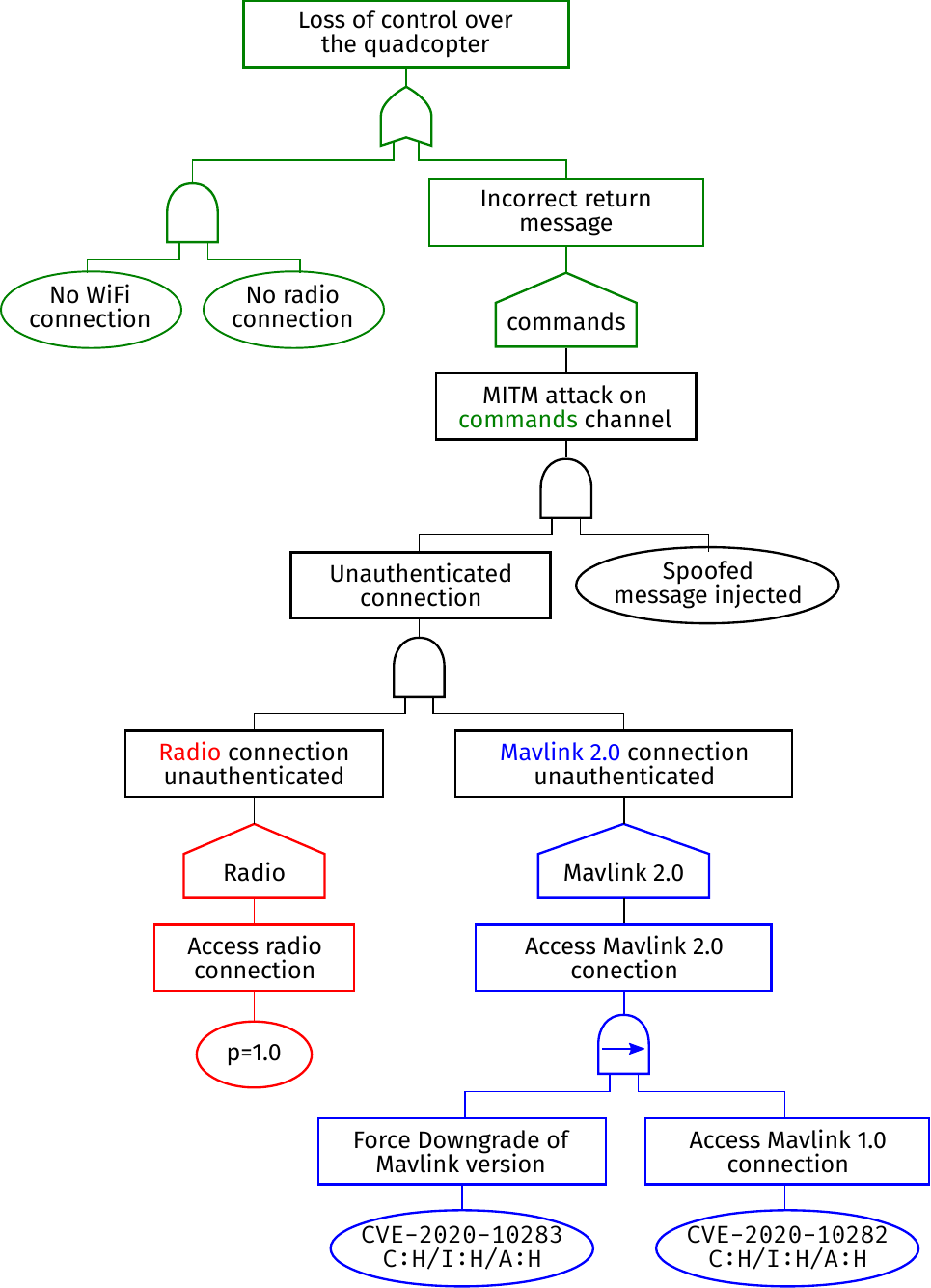}
    \caption{Combined AFT of the example.}
    \label{fig:aft}
    \vspace{-0.8cm}
\end{figure}

Similarly, combination rules can express the spreading of faults along data flow or how an attack on a vulnerable library allows the attacker to also compromise a component using it -- thereby connecting a logical component from the Dataflow Graph with the libraries its realization uses.

In our example scenario, the Fault Tree leading to control loss and theft of the drone (Figure~\ref{fig:faultTree}) and the Attack Tree in Figure~\ref{fig:attackt} can be combined into the AFT shown in Figure~\ref{fig:aft}. The Dataflow Graph and Deployment Model---in the transient reconfiguration state, when the WiFi connection is lost, but the return signal is not yet sent---provide the necessary information to join the models despite their incompatible abstraction levels. Two combination rules are used to combine the trees. These are derived from two CAPEC patterns:

\emph{CAPEC-115: Authentication Bypass} A channel is accessible without authentication if all constituent protocols, hardware, etc. is accessible without authentication.

\emph{CAPEC-594: Traffic Injection} A malicious actor can insert a spoofed message into a channel, if it requires no authentication.

These rules are used to generate the partial tree (black) that connects the external events of the Attack and Fault Trees. The first rule matches the deployment of the \emph{commands} channel, connecting it to attacks on the authentication of the platforms and protocols used. Next, the second rule matches the \emph{commands} channel in the dataflow and is connected to the external event in the fault tree.

The resulting AFT shows, that the Mavlink vulnerability in combination with using radio communication can lead to the control loss over the quadcopter. This state occurs during adaptation, after the quadcopter flies out of WiFi range by injecting a false return message into the commands channel.


\section{Analysis Approach}
\label{sec:analysis}
In this section, we provide an outlook on the two dimensions of our planned analysis approach.
\paragraph{Attack-Fault Tree Analysis} For each state of the system (transient as well as steady), the corresponding combined AFT is generated. Existing approaches for analyzing AFTs can be used, e.g. first translating the AFT into stochastic timed automata as input to existing model checkers, e.\,g.\ Uppaal SMC \cite{7911867}.
Our proposed combination approach for ATs and FTs will be extended to provide the necessary timing and probability information. Already, events in ATs and FTs are often annotated with timing or probability information. By enriching the Dataflow Graph with additional timing information, representing the delay and frequency of messages or the processing time of these messages in components, timing constraints on adaptations can be derived in the analysis, similar to \cite{Priesterjahn2013}. For example, an adaptation to avoid the spread of an error across multiple components must be faster than the spreading of the error.

\paragraph{Henshin state space analysis} Since we model the individual steps of an adaptation using Henshin transformation rules, we can also use the state space analysis provided by the Henshin tool set~\cite{10.1007/978-3-642-16145-2_9}.
This allows to explore the entire state space resulting from the transformations, including the transient states that may occur during adaptation with respect to the Dataflow Graph and the Deployment Model.
By examining the state space, resulting reachable states can subsequently be examined in more detail with the help of the AFT.
This requires applying the individual transformations to the models and then generating a new AFT.
In addition, the correctness of the modeling of the adaptation can be examined with the help of invariants. 
For example, to ensure that there exists no combination of transformations that results in the command channel communicating via radio and WiFi at the same time. 


\section{Conclusion and Future Work}\label{sec:conclusion}
In this vision paper, we present an integrated approach to model and analyze safety and security aspects of a self-adaptive system. Bridging the gap between high-level modeling and low-level implementation on specific platforms enables the analysis of vulnerabilities of system components with respect to safety requirements. We also consider intermediate steps of an adaptation phase, which may provide new attack surfaces. The next steps of our research are to implement the different models and to define precisely the translation into AFTs. 
Currently, our approach does not consider time and probability aspects, although this is an important point since, e.g., long lasting high effort attacks on vulnerabilities that exist for only a short time during adaptation are less likely than attacks that can be executed within a short time~\cite{10.1007/978-3-319-22975-1_11,Stoelinga2}. 
In addition, do we plan to use our approach also at run-time to continuously analyze a system with respect to safety and security.
This allows us to include emerging security vulnerabilities even during the runtime of the system.
To do this, however, we must ensure that the models are kept in sync with the system state as efficiently as possible and that the translation and analysis of the AFT can be performed within fixed time limits~\cite{szvetits2016systematic,Bennaceur2014}. 
Additionally, we plan a case study in which we use our approach to analyze larger and realistic systems in order to evaluate its scalability. 


\begin{acks}
This work was partially supported by the Austrian Science Fund (FWF): I 4701-N and German Research Foundation (DFG): 435878599.
\end{acks}

\balance
\bibliographystyle{ACM-Reference-Format}
\bibliography{references}


\begin{thebibliography}{30}


\ifx \showCODEN    \undefined \def \showCODEN     #1{\unskip}     \fi
\ifx \showDOI      \undefined \def \showDOI       #1{#1}\fi
\ifx \showISBNx    \undefined \def \showISBNx     #1{\unskip}     \fi
\ifx \showISBNxiii \undefined \def \showISBNxiii  #1{\unskip}     \fi
\ifx \showISSN     \undefined \def \showISSN      #1{\unskip}     \fi
\ifx \showLCCN     \undefined \def \showLCCN      #1{\unskip}     \fi
\ifx \shownote     \undefined \def \shownote      #1{#1}          \fi
\ifx \showarticletitle \undefined \def \showarticletitle #1{#1}   \fi
\ifx \showURL      \undefined \def \showURL       {\relax}        \fi
\providecommand\bibfield[2]{#2}
\providecommand\bibinfo[2]{#2}
\providecommand\natexlab[1]{#1}
\providecommand\showeprint[2][]{arXiv:#2}

\bibitem[Andr{\'{e}} et~al\mbox{.}(2019)]%
        {Stoelinga2}
\bibfield{author}{\bibinfo{person}{{\'{E}}tienne Andr{\'{e}}}, \bibinfo{person}{Didier Lime}, \bibinfo{person}{Mathias Ramparison}, {and} \bibinfo{person}{Mari{\"{e}}lle Stoelinga}.} \bibinfo{year}{2019}\natexlab{}.
\newblock \showarticletitle{Parametric analyses of attack-fault trees}.
\newblock \bibinfo{journal}{\emph{CoRR}}  \bibinfo{volume}{abs/1902.04336} (\bibinfo{year}{2019}).
\newblock
\showeprint[arXiv]{1902.04336}
\urldef\tempurl%
\url{http://arxiv.org/abs/1902.04336}
\showURL{%
\tempurl}


\bibitem[Arendt et~al\mbox{.}(2010)]%
        {10.1007/978-3-642-16145-2_9}
\bibfield{author}{\bibinfo{person}{Thorsten Arendt}, \bibinfo{person}{Enrico Biermann}, \bibinfo{person}{Stefan Jurack}, \bibinfo{person}{Christian Krause}, {and} \bibinfo{person}{Gabriele Taentzer}.} \bibinfo{year}{2010}\natexlab{}.
\newblock \showarticletitle{Henshin: Advanced Concepts and Tools for In-Place EMF Model Transformations}. In \bibinfo{booktitle}{\emph{Model Driven Engineering Languages and Systems}}, \bibfield{editor}{\bibinfo{person}{Dorina~C. Petriu}, \bibinfo{person}{Nicolas Rouquette}, {and} \bibinfo{person}{{\O}ystein Haugen}} (Eds.). \bibinfo{publisher}{Springer Berlin Heidelberg}, \bibinfo{address}{Berlin, Heidelberg}, \bibinfo{pages}{121--135}.
\newblock
\showISBNx{978-3-642-16145-2}
\urldef\tempurl%
\url{https://doi.org/10.1007/978-3-642-16145-2_9}
\showDOI{\tempurl}


\bibitem[Bennaceur et~al\mbox{.}(2014)]%
        {Bennaceur2014}
\bibfield{author}{\bibinfo{person}{Amel Bennaceur}, \bibinfo{person}{Robert France}, \bibinfo{person}{Giordano Tamburrelli}, \bibinfo{person}{Thomas Vogel}, \bibinfo{person}{Pieter~J. Mosterman}, \bibinfo{person}{Walter Cazzola}, \bibinfo{person}{Fabio~M. Costa}, \bibinfo{person}{Alfonso Pierantonio}, \bibinfo{person}{Matthias Tichy}, \bibinfo{person}{Mehmet Ak{\c{s}}it}, \bibinfo{person}{P{\"a}r Emmanuelson}, \bibinfo{person}{Huang Gang}, \bibinfo{person}{Nikolaos Georgantas}, {and} \bibinfo{person}{David Redlich}.} \bibinfo{year}{2014}\natexlab{}.
\newblock \bibinfo{booktitle}{\emph{Mechanisms for Leveraging Models at Runtime in Self-adaptive Software}}.
\newblock \bibinfo{publisher}{Springer International Publishing}, \bibinfo{address}{Cham}, \bibinfo{pages}{19--46}.
\newblock
\showISBNx{978-3-319-08915-7}
\urldef\tempurl%
\url{https://doi.org/10.1007/978-3-319-08915-7_2}
\showDOI{\tempurl}


\bibitem[Bucchiarone et~al\mbox{.}(2015)]%
        {Bucchiarone2015}
\bibfield{author}{\bibinfo{person}{Antonio Bucchiarone}, \bibinfo{person}{Hartmut Ehrig}, \bibinfo{person}{Claudia Ermel}, \bibinfo{person}{Patrizio Pelliccione}, {and} \bibinfo{person}{Olga Runge}.} \bibinfo{year}{2015}\natexlab{}.
\newblock \bibinfo{booktitle}{\emph{Rule-Based Modeling and Static Analysis of Self-adaptive Systems by Graph Transformation}}.
\newblock \bibinfo{publisher}{Springer International Publishing}, \bibinfo{address}{Cham}, \bibinfo{pages}{582--601}.
\newblock
\showISBNx{978-3-319-15545-6}
\urldef\tempurl%
\url{https://doi.org/10.1007/978-3-319-15545-6_33}
\showDOI{\tempurl}


\bibitem[Camilli et~al\mbox{.}(2021)]%
        {Scandurra2021}
\bibfield{author}{\bibinfo{person}{Matteo Camilli}, \bibinfo{person}{Raffaela Mirandola}, {and} \bibinfo{person}{Patrizia Scandurra}.} \bibinfo{year}{2021}\natexlab{}.
\newblock \showarticletitle{Runtime Equilibrium Verification for Resilient Cyber-Physical Systems}. In \bibinfo{booktitle}{\emph{2021 IEEE International Conference on Autonomic Computing and Self-Organizing Systems (ACSOS)}}. \bibinfo{pages}{71--80}.
\newblock
\urldef\tempurl%
\url{https://doi.org/10.1109/ACSOS52086.2021.00025}
\showDOI{\tempurl}


\bibitem[Causevic et~al\mbox{.}(2019)]%
        {8753949}
\bibfield{author}{\bibinfo{person}{Aida Causevic}, \bibinfo{person}{Alessandro~V. Papadopoulos}, {and} \bibinfo{person}{Marjan Sirjani}.} \bibinfo{year}{2019}\natexlab{}.
\newblock \showarticletitle{Towards a Framework for Safe and Secure Adaptive Collaborative Systems}. In \bibinfo{booktitle}{\emph{2019 IEEE 43rd Annual Computer Software and Applications Conference (COMPSAC)}}, Vol.~\bibinfo{volume}{2}. \bibinfo{pages}{165--170}.
\newblock
\urldef\tempurl%
\url{https://doi.org/10.1109/COMPSAC.2019.10201}
\showDOI{\tempurl}


\bibitem[Khakpour et~al\mbox{.}(2019)]%
        {Khakpour2019}
\bibfield{author}{\bibinfo{person}{Narges Khakpour}, \bibinfo{person}{Charilaos Skandylas}, \bibinfo{person}{Goran~Saman Nariman}, {and} \bibinfo{person}{Danny Weyns}.} \bibinfo{year}{2019}\natexlab{}.
\newblock \showarticletitle{Towards Secure Architecture-Based Adaptations}. In \bibinfo{booktitle}{\emph{2019 IEEE/ACM 14th International Symposium on Software Engineering for Adaptive and Self-Managing Systems (SEAMS)}}. \bibinfo{pages}{114--125}.
\newblock
\urldef\tempurl%
\url{https://doi.org/10.1109/SEAMS.2019.00023}
\showDOI{\tempurl}


\bibitem[Klumpp et~al\mbox{.}(2016)]%
        {Klumpp2016}
\bibfield{author}{\bibinfo{person}{Dominik Klumpp}, \bibinfo{person}{Axel Habermaier}, \bibinfo{person}{Benedikt Eberhardinger}, {and} \bibinfo{person}{Hella Seebach}.} \bibinfo{year}{2016}\natexlab{}.
\newblock \showarticletitle{Optimising Runtime Safety Analysis Efficiency for Self-Organising Systems}. In \bibinfo{booktitle}{\emph{2016 IEEE 1st International Workshops on Foundations and Applications of Self* Systems (FAS*W)}}. \bibinfo{pages}{120--125}.
\newblock
\urldef\tempurl%
\url{https://doi.org/10.1109/FAS-W.2016.37}
\showDOI{\tempurl}


\bibitem[Kotenko and Chechulin(2013)]%
        {6568374}
\bibfield{author}{\bibinfo{person}{Igor Kotenko} {and} \bibinfo{person}{Andrey Chechulin}.} \bibinfo{year}{2013}\natexlab{}.
\newblock \showarticletitle{A Cyber Attack Modeling and Impact Assessment framework}. In \bibinfo{booktitle}{\emph{2013 5th International Conference on Cyber Conflict (CYCON 2013)}}. \bibinfo{pages}{1--24}.
\newblock


\bibitem[Kumar et~al\mbox{.}(2015)]%
        {10.1007/978-3-319-22975-1_11}
\bibfield{author}{\bibinfo{person}{Rajesh Kumar}, \bibinfo{person}{Enno Ruijters}, {and} \bibinfo{person}{Mari{\"e}lle Stoelinga}.} \bibinfo{year}{2015}\natexlab{}.
\newblock \showarticletitle{Quantitative Attack Tree Analysis via Priced Timed Automata}. In \bibinfo{booktitle}{\emph{Formal Modeling and Analysis of Timed Systems}}, \bibfield{editor}{\bibinfo{person}{Sriram Sankaranarayanan} {and} \bibinfo{person}{Enrico Vicario}} (Eds.). \bibinfo{publisher}{Springer International Publishing}, \bibinfo{address}{Cham}, \bibinfo{pages}{156--171}.
\newblock
\showISBNx{978-3-319-22975-1}
\urldef\tempurl%
\url{https://doi.org/10.1007/978-3-319-22975-1_11}
\showDOI{\tempurl}


\bibitem[Kumar and Stoelinga(2017)]%
        {7911867}
\bibfield{author}{\bibinfo{person}{Rajesh Kumar} {and} \bibinfo{person}{Mariëlle Stoelinga}.} \bibinfo{year}{2017}\natexlab{}.
\newblock \showarticletitle{Quantitative Security and Safety Analysis with Attack-Fault Trees}. In \bibinfo{booktitle}{\emph{2017 IEEE 18th International Symposium on High Assurance Systems Engineering (HASE)}}. \bibinfo{pages}{25--32}.
\newblock
\urldef\tempurl%
\url{https://doi.org/10.1109/HASE.2017.12}
\showDOI{\tempurl}


\bibitem[Liu et~al\mbox{.}(2017)]%
        {10.1145/3055186.3055193}
\bibfield{author}{\bibinfo{person}{Jiafa Liu}, \bibinfo{person}{Di Ma}, \bibinfo{person}{Andre Weimerskirch}, {and} \bibinfo{person}{Haojin Zhu}.} \bibinfo{year}{2017}\natexlab{}.
\newblock \showarticletitle{A Functional Co-Design towards Safe and Secure Vehicle Platooning}. In \bibinfo{booktitle}{\emph{Proceedings of the 3rd ACM Workshop on Cyber-Physical System Security}} (Abu Dhabi, United Arab Emirates) \emph{(\bibinfo{series}{CPSS '17})}. \bibinfo{publisher}{Association for Computing Machinery}, \bibinfo{address}{New York, NY, USA}, \bibinfo{pages}{81–90}.
\newblock
\showISBNx{9781450349567}
\urldef\tempurl%
\url{https://doi.org/10.1145/3055186.3055193}
\showDOI{\tempurl}


\bibitem[{Nai Fovino} et~al\mbox{.}(2009a)]%
        {NAIFOVINO20091394}
\bibfield{author}{\bibinfo{person}{Igor {Nai Fovino}}, \bibinfo{person}{Marcelo Masera}, {and} \bibinfo{person}{Alessio {De Cian}}.} \bibinfo{year}{2009}\natexlab{a}.
\newblock \showarticletitle{Integrating cyber attacks within fault trees}.
\newblock \bibinfo{journal}{\emph{Reliability Engineering \& System Safety}} \bibinfo{volume}{94}, \bibinfo{number}{9} (\bibinfo{year}{2009}), \bibinfo{pages}{1394--1402}.
\newblock
\showISSN{0951-8320}
\urldef\tempurl%
\url{https://doi.org/10.1016/j.ress.2009.02.020}
\showDOI{\tempurl}
\newblock
\shownote{ESREL 2007, the 18th European Safety and Reliability Conference}.


\bibitem[{Nai Fovino} et~al\mbox{.}(2009b)]%
        {Fovino2009}
\bibfield{author}{\bibinfo{person}{Igor {Nai Fovino}}, \bibinfo{person}{Marcelo Masera}, {and} \bibinfo{person}{Alessio {De Cian}}.} \bibinfo{year}{2009}\natexlab{b}.
\newblock \showarticletitle{Integrating cyber attacks within fault trees}.
\newblock \bibinfo{journal}{\emph{Reliability Engineering \& System Safety}} \bibinfo{volume}{94}, \bibinfo{number}{9} (\bibinfo{year}{2009}), \bibinfo{pages}{1394--1402}.
\newblock
\showISSN{0951-8320}
\urldef\tempurl%
\url{https://doi.org/10.1016/j.ress.2009.02.020}
\showDOI{\tempurl}
\newblock
\shownote{ESREL 2007, the 18th European Safety and Reliability Conference}.


\bibitem[Ou et~al\mbox{.}(2006)]%
        {Ou2006}
\bibfield{author}{\bibinfo{person}{Xinming Ou}, \bibinfo{person}{Wayne~F. Boyer}, {and} \bibinfo{person}{Miles~A. McQueen}.} \bibinfo{year}{2006}\natexlab{}.
\newblock \showarticletitle{A Scalable Approach to Attack Graph Generation}. In \bibinfo{booktitle}{\emph{Proceedings of the 13th ACM Conference on Computer and Communications Security}} (Alexandria, Virginia, USA) \emph{(\bibinfo{series}{CCS '06})}. \bibinfo{publisher}{Association for Computing Machinery}, \bibinfo{address}{New York, NY, USA}, \bibinfo{pages}{336–345}.
\newblock
\showISBNx{1595935185}
\urldef\tempurl%
\url{https://doi.org/10.1145/1180405.1180446}
\showDOI{\tempurl}


\bibitem[Ou et~al\mbox{.}(2005)]%
        {MulVAL2005}
\bibfield{author}{\bibinfo{person}{Xinming Ou}, \bibinfo{person}{Sudhakar Govindavajhala}, {and} \bibinfo{person}{Andrew~W. Appel}.} \bibinfo{year}{2005}\natexlab{}.
\newblock \showarticletitle{MulVAL: A Logic-Based Network Security Analyzer}. In \bibinfo{booktitle}{\emph{Proceedings of the 14th Conference on USENIX Security Symposium - Volume 14}} (Baltimore, MD) \emph{(\bibinfo{series}{SSYM'05})}. \bibinfo{publisher}{USENIX Association}, \bibinfo{address}{USA}, \bibinfo{pages}{8}.
\newblock


\bibitem[Pacheco and Hariri(2016)]%
        {IDS2}
\bibfield{author}{\bibinfo{person}{Jesus Pacheco} {and} \bibinfo{person}{Salim Hariri}.} \bibinfo{year}{2016}\natexlab{}.
\newblock \showarticletitle{IoT Security Framework for Smart Cyber Infrastructures}. In \bibinfo{booktitle}{\emph{2016 IEEE 1st International Workshops on Foundations and Applications of Self* Systems (FAS*W)}}. \bibinfo{pages}{242--247}.
\newblock
\urldef\tempurl%
\url{https://doi.org/10.1109/FAS-W.2016.58}
\showDOI{\tempurl}


\bibitem[Pacheco et~al\mbox{.}(2017)]%
        {IDS1}
\bibfield{author}{\bibinfo{person}{Jesus Pacheco}, \bibinfo{person}{Xiaoyang Zhu}, \bibinfo{person}{Youakim Badr}, {and} \bibinfo{person}{Salim Hariri}.} \bibinfo{year}{2017}\natexlab{}.
\newblock \showarticletitle{Enabling Risk Management for Smart Infrastructures with an Anomaly Behavior Analysis Intrusion Detection System}. In \bibinfo{booktitle}{\emph{2017 IEEE 2nd International Workshops on Foundations and Applications of Self* Systems (FAS*W)}}. \bibinfo{pages}{324--328}.
\newblock
\urldef\tempurl%
\url{https://doi.org/10.1109/FAS-W.2017.167}
\showDOI{\tempurl}


\bibitem[Pianini et~al\mbox{.}(2019)]%
        {Pianini2019}
\bibfield{author}{\bibinfo{person}{Danilo Pianini}, \bibinfo{person}{Roberto Casadei}, {and} \bibinfo{person}{Mirko Viroli}.} \bibinfo{year}{2019}\natexlab{}.
\newblock \showarticletitle{Security in Collective Adaptive Systems: A Roadmap}. In \bibinfo{booktitle}{\emph{2019 IEEE 4th International Workshops on Foundations and Applications of Self* Systems (FAS*W)}}. \bibinfo{pages}{86--91}.
\newblock
\urldef\tempurl%
\url{https://doi.org/10.1109/FAS-W.2019.00034}
\showDOI{\tempurl}


\bibitem[Priesterjahn et~al\mbox{.}(2013)]%
        {Priesterjahn2013}
\bibfield{author}{\bibinfo{person}{Claudia Priesterjahn}, \bibinfo{person}{Dominik Steenken}, {and} \bibinfo{person}{Matthias Tichy}.} \bibinfo{year}{2013}\natexlab{}.
\newblock \bibinfo{booktitle}{\emph{Timed Hazard Analysis of Self-healing Systems}}.
\newblock \bibinfo{publisher}{Springer Berlin Heidelberg}, \bibinfo{address}{Berlin, Heidelberg}, \bibinfo{pages}{112--151}.
\newblock
\showISBNx{978-3-642-36249-1}
\urldef\tempurl%
\url{https://doi.org/10.1007/978-3-642-36249-1_5}
\showDOI{\tempurl}


\bibitem[Quigley et~al\mbox{.}(2009)]%
        {quigley2009ros}
\bibfield{author}{\bibinfo{person}{Morgan Quigley}, \bibinfo{person}{Ken Conley}, \bibinfo{person}{Brian Gerkey}, \bibinfo{person}{Josh Faust}, \bibinfo{person}{Tully Foote}, \bibinfo{person}{Jeremy Leibs}, \bibinfo{person}{Rob Wheeler}, \bibinfo{person}{Andrew~Y Ng}, {et~al\mbox{.}}} \bibinfo{year}{2009}\natexlab{}.
\newblock \showarticletitle{ROS: an open-source Robot Operating System}. In \bibinfo{booktitle}{\emph{ICRA workshop on open source software}}, Vol.~\bibinfo{volume}{3}. Kobe, Japan, \bibinfo{pages}{5}.
\newblock


\bibitem[Ruijters and Stoelinga(2015)]%
        {RUIJTERS201529}
\bibfield{author}{\bibinfo{person}{Enno Ruijters} {and} \bibinfo{person}{Mariëlle Stoelinga}.} \bibinfo{year}{2015}\natexlab{}.
\newblock \showarticletitle{Fault tree analysis: A survey of the state-of-the-art in modeling, analysis and tools}.
\newblock \bibinfo{journal}{\emph{Computer Science Review}}  \bibinfo{volume}{15-16} (\bibinfo{year}{2015}), \bibinfo{pages}{29--62}.
\newblock
\showISSN{1574-0137}
\urldef\tempurl%
\url{https://doi.org/10.1016/j.cosrev.2015.03.001}
\showDOI{\tempurl}


\bibitem[Schneier(2015)]%
        {schneier2015secrets}
\bibfield{author}{\bibinfo{person}{Bruce Schneier}.} \bibinfo{year}{2015}\natexlab{}.
\newblock \bibinfo{booktitle}{\emph{Secrets and Lies: Digital Security in a Networked World}}.
\newblock \bibinfo{publisher}{Wiley}.
\newblock
\showISBNx{9781119092438}
\showLCCN{2015932613}


\bibitem[Scoccia et~al\mbox{.}(2020)]%
        {ACSThreatDatabase}
\bibfield{author}{\bibinfo{person}{Gian~Luca Scoccia}, \bibinfo{person}{Marco Autili}, {and} \bibinfo{person}{Paola Inverardi}.} \bibinfo{year}{2020}\natexlab{}.
\newblock \showarticletitle{A self-configuring and adaptive privacy-aware permission system for Android apps}. In \bibinfo{booktitle}{\emph{2020 IEEE International Conference on Autonomic Computing and Self-Organizing Systems (ACSOS)}}. \bibinfo{pages}{38--47}.
\newblock
\urldef\tempurl%
\url{https://doi.org/10.1109/ACSOS49614.2020.00024}
\showDOI{\tempurl}


\bibitem[Steiner and Liggesmeyer(2013)]%
        {SteinerLiggesmeyer2016}
\bibfield{author}{\bibinfo{person}{Max Steiner} {and} \bibinfo{person}{Peter Liggesmeyer}.} \bibinfo{year}{2013}\natexlab{}.
\newblock \showarticletitle{Combination of Safety and Security Analysis - Finding Security Problems That Threaten the Safety of a System}. In \bibinfo{booktitle}{\emph{32nd International Conference on Computer Safety, Reliability and Security (SAFECOMP 2013). Workshops and Tutorials : CARS, SASSUR, DECS, ASCOMS}}.
\newblock
\urldef\tempurl%
\url{http://nbn-resolving.de/urn:nbn:de:hbz:386-kluedo-43604}
\showURL{%
\tempurl}


\bibitem[Str{\"u}ber et~al\mbox{.}(2017)]%
        {10.1007/978-3-319-61470-0_12}
\bibfield{author}{\bibinfo{person}{Daniel Str{\"u}ber}, \bibinfo{person}{Kristopher Born}, \bibinfo{person}{Kanwal~Daud Gill}, \bibinfo{person}{Raffaela Groner}, \bibinfo{person}{Timo Kehrer}, \bibinfo{person}{Manuel Ohrndorf}, {and} \bibinfo{person}{Matthias Tichy}.} \bibinfo{year}{2017}\natexlab{}.
\newblock \showarticletitle{Henshin: A Usability-Focused Framework for EMF Model Transformation Development}. In \bibinfo{booktitle}{\emph{Graph Transformation}}, \bibfield{editor}{\bibinfo{person}{Juan de~Lara} {and} \bibinfo{person}{Detlef Plump}} (Eds.). \bibinfo{publisher}{Springer International Publishing}, \bibinfo{address}{Cham}, \bibinfo{pages}{196--208}.
\newblock
\showISBNx{978-3-319-61470-0}
\urldef\tempurl%
\url{https://doi.org/10.1007/978-3-319-61470-0_12}
\showDOI{\tempurl}


\bibitem[Szvetits and Zdun(2016)]%
        {szvetits2016systematic}
\bibfield{author}{\bibinfo{person}{Michael Szvetits} {and} \bibinfo{person}{Uwe Zdun}.} \bibinfo{year}{2016}\natexlab{}.
\newblock \showarticletitle{Systematic literature review of the objectives, techniques, kinds, and architectures of models at runtime}.
\newblock \bibinfo{journal}{\emph{Software \& Systems Modeling}} \bibinfo{volume}{15}, \bibinfo{number}{1} (\bibinfo{year}{2016}), \bibinfo{pages}{31--69}.
\newblock
\urldef\tempurl%
\url{https://doi.org/10.1007/s10270-013-0394-9}
\showDOI{\tempurl}


\bibitem[Veledar et~al\mbox{.}(2019)]%
        {S3}
\bibfield{author}{\bibinfo{person}{Omar Veledar}, \bibinfo{person}{Violeta Damjanovic-Behrendt}, {and} \bibinfo{person}{Georg Macher}.} \bibinfo{year}{2019}\natexlab{}.
\newblock \showarticletitle{{{{Digital Twins for Dependability Improvement of Autonomous Driving}}}}. In \bibinfo{booktitle}{\emph{Systems, Software and Services Process Improvement}}, \bibfield{editor}{\bibinfo{person}{Alastair Walker}, \bibinfo{person}{Rory~V. O'Connor}, {and} \bibinfo{person}{Richard Messnarz}} (Eds.). \bibinfo{publisher}{Springer International Publishing}, \bibinfo{address}{Cham}, \bibinfo{pages}{415--426}.
\newblock
\showISBNx{978-3-030-28005-5}
\urldef\tempurl%
\url{https://doi.org/10.1007/978-3-030-28005-5_32}
\showDOI{\tempurl}


\bibitem[Vesely et~al\mbox{.}(1981)]%
        {vesely1981fault}
\bibfield{author}{\bibinfo{person}{William~E Vesely}, \bibinfo{person}{Francine~F Goldberg}, \bibinfo{person}{Norman~H Roberts}, {and} \bibinfo{person}{David~F Haasl}.} \bibinfo{year}{1981}\natexlab{}.
\newblock \bibinfo{booktitle}{\emph{Fault tree handbook}}.
\newblock \bibinfo{type}{{T}echnical {R}eport}. \bibinfo{institution}{Nuclear Regulatory Commission Washington DC}.
\newblock


\bibitem[Viertel et~al\mbox{.}(2019)]%
        {Viertel2019}
\bibfield{author}{\bibinfo{person}{Fabien~Patrick Viertel}, \bibinfo{person}{Fabian Kortum}, \bibinfo{person}{Leif Wagner}, {and} \bibinfo{person}{Kurt Schneider}.} \bibinfo{year}{2019}\natexlab{}.
\newblock \showarticletitle{Are Third-Party Libraries Secure? A Software Library Checker for Java}. In \bibinfo{booktitle}{\emph{Risks and Security of Internet and Systems}}, \bibfield{editor}{\bibinfo{person}{Akka Zemmari}, \bibinfo{person}{Mohamed Mosbah}, \bibinfo{person}{Nora Cuppens-Boulahia}, {and} \bibinfo{person}{Fr{\'e}d{\'e}ric Cuppens}} (Eds.). \bibinfo{publisher}{Springer International Publishing}, \bibinfo{address}{Cham}, \bibinfo{pages}{18--34}.
\newblock
\showISBNx{978-3-030-12143-3}


\end{thebibliography}



\end{document}